\documentclass[aps,prb,reprint,showpacs,superscriptaddress,amsmath,amssymb]{revtex4-1}

\usepackage{graphicx}
\usepackage{dcolumn} 
\usepackage{bm} 
\usepackage{hyperref} 
\usepackage{color}

\definecolor{red}{rgb}{0,0,0}
\newcommand{\red}{\color{red}}
\definecolor{green}{rgb}{0,1,0}

\begin{document}

\title{One-dimensional quantum antiferromagnetism in the $p-$orbital CsO$_2$ compound revealed by electron paramagnetic resonance}
\author{Tilen Knafli\v{c}}
\affiliation{Jo\v{z}ef Stefan Institute, Jamova 39, SI-1000 Ljubljana, Slovenia}
\author{Martin Klanj\v{s}ek}
\affiliation{ Jo\v{z}ef Stefan Institute, Jamova 39, SI-1000 Ljubljana, Slovenia}
\author{Annette Sans}
\affiliation{Max-Planck-Institut fur Chemische Physik fester Stoffe, 01187 Dresden, Germany}
\author{Peter Adler}
\affiliation{Max-Planck-Institut fur Chemische Physik fester Stoffe, 01187 Dresden, Germany}
\author{Martin Jansen}
\affiliation{Max-Planck-Institut fur Chemische Physik fester Stoffe, 01187 Dresden, Germany}
\affiliation{Max-Planck-Institut fur Festk\"{o}rperforschung, 70569 Stuttgart, Germany}
\author{Claudia Felser}
\affiliation{Max-Planck-Institut fur Chemische Physik fester Stoffe, 01187 Dresden, Germany}
\author{Denis Ar\v{c}on}
\email{denis.arcon@ijs.si}
\affiliation{ Jo\v{z}ef Stefan Institute, Jamova 39, SI-1000 Ljubljana, Slovenia}
\affiliation{Faculty of Mathematics and Physics, University of Ljubljana, Jadranska 19, SI-1000 Ljubljana, Slovenia}

\begin{abstract}
Recently it was proposed that the orbital ordering of $\pi_{x,y}^*$ molecular orbitals in the superoxide CsO$_2$ compound leads to the formation of spin-1/2 chains below the structural phase transition occuring at $T_{\rm{s1}}=61$~K on cooling. Here we report a detailed X-band electron paramagnetic resonance (EPR) study of this phase in CsO$_2$ powder. The EPR signal appears as a broad line below  $T_{\rm{s1}}$, which is replaced by the antiferromagnetic resonance below the N\'{e}el temperature $T_{\rm N}=8.3$~K. The temperature dependence of the EPR linewidth between $T_{\rm{s1}}$  and $T_{\rm{N}}$  agrees with the predictions for the one-dimensional Heisenberg antiferromagnetic chain of $S=1/2$ spins in the presence of symmetric anisotropic exchange interaction. Complementary analysis of the EPR lineshape, linewidth and the signal intensity within the Tomonaga-Luttinger liquid (TLL) framework allows for a determination of the TLL exponent $K=0.48$. Present EPR data  thus fully comply with the quantum antiferromagnetic state of spin-1/2 chains in the orbitally ordered phase of CsO$_2$, which is, therefore, a unique $p-$orbital system where such a state could be studied.
\end{abstract}

\pacs{75.10.Pq, 71.10.Pm, 76.30.-v}

\maketitle

\section{Introduction}
Quantum antiferromagnets have provided one of the most fertile grounds for discoveries of  new states of matter and testing the theoretical models. 
The one-dimensional Heisenberg antiferromagnetic chain of $S=1/2$ spins (1D-HAF) is probably the most studied example of such quantum antiferromagnets where recent advances in the theoretical approaches\cite{Giamarchi, O-A1, O-A2, Schon} have enabled quantitative discussion of experimental data on model systems. 
Among the best realizations of 1D-HAFs are KCuF$_3$,\cite{Lake, Lake2, Chakh,Loidl} Sr$_2$CuO$_3$,\cite{Motoyama,Takigawa, Takigawa2} CuSO$_4\cdot 5$D$_2$O,\cite{Mourigal}  Cu(C$_4$H$_4$N$_2$)(NO$_3$)$_2$,\cite{Hammar, Tom, Kono}  copper-pyrimidine-dinitrate\cite{Zvyagin} and CuSe$_2$O$_5$.\cite{Rosner, Mirta, Mirta2} Excellent agreement between the extensive experimental and theoretical work established the existence of  gapless $S = 1/2$ excitations, the so called spinons,\cite{Faddeev} into which the conventional $S = 1$ magnon excitations fractionalize.  The low-energy description of gapless phases in 1D-HAFs leads to the Tomonaga-Luttinger liquid (TLL) paradigm\cite{Giamarchi} as a universal concept of quantum physics in one dimension, beyond the perturbation theory.

\begin{figure}[htbp]
  \includegraphics[width=\linewidth]{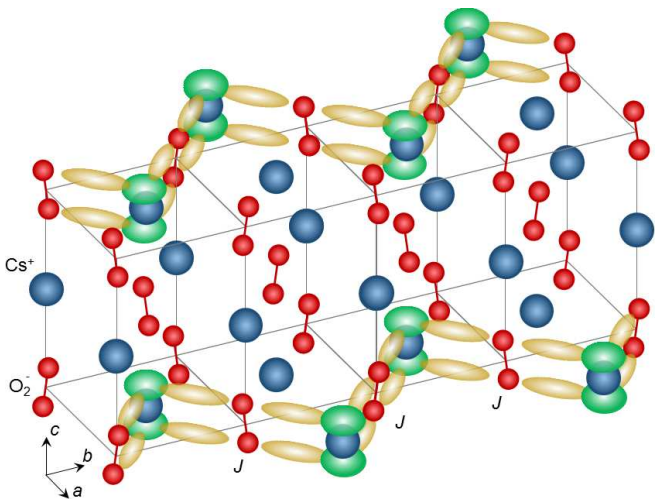}
  \caption[]
   {(color online) {\red Sketch of the CsO$_2$ crystal structure and relevant orbitals.} Zig-zag chains of  ordered O$_2^-$ $\pi_{x,y}^{*}$ molecular orbitals ($S=1/2$) run along the crystallographic $b-$axis. Blue and red spheres represent caesium and oxygen atoms, respectively. The superexchange interactions, $J$, between $\pi_{x,y}^{*}$ orbitals (yellow) are bridged by Cs $p_z$ orbitals (green).}
   \label{fig_struct}
\end{figure}

In all the above-cited realizations of 1D-HAFs, the  spin states stem from the electrons in $d$orbitals of transition metal ions. Recently, an important structural and magnetic susceptibility investigation of  CsO$_2$, a member of the alkali superoxide {\em A}O$_2$ ({\em A} = alkali metal) family,\cite{Jansen_rev, Sarma,Solo,Kim,Solo2,Oles,Palstra,Palstra2,Martin_CsO2,Koenzig} suggested this compound to realize the first $p-$orbital 1D-HAF.\cite{Palstra} 1D-HAF state is established in the orbitally-ordered phase below $T_{\rm{s1}}=61$~K on cooling, 
where a spatial alignment of the $\pi_{x,y}^*$ molecular orbitals of O$_2^{-}$ units with $S=1/2$ results in a strong superexchange interaction $J$ along the crystallographic $b-$axis, i.e., the chain axis, while in the perpendicular crystallographic $a-$direction the
exchange interaction is very small due to poor overlap of the corresponding molecular  orbitals (Fig. \ref{fig_struct}).
Comprehensive $^{133}$Cs nuclear magnetic resonance (NMR) measurements of spin-lattice relaxation rate $T_1^{-1}$ in different magnetic fields indeed revealed the characteristic power law dependence $T_1^{-1}\propto T^\nu$,\cite{Martin_CsO2} which is a hallmark of the TLL state.\cite{Mladen, Martin_BPCB} The field-dependence of the extracted  parameters quantitatively demonstrates that the low-energy spectrum of spin-fluctuations in the 1D-HAF state of CsO$_2$ can be described in terms of TLL yielding a dimensionless TLL exponent $K\approx 1/4$, consistent with the Ising-like exchange-coupling anisotropy. 
Because the proposed 1D-HAF state in CsO$_2$ is based on $p-$orbital states that occupy a different regime of interplay between magnetism, structure, vibrations, and orbital ordering than the more conventional $d-$orbital or $f-$orbital states, 
finding additional independent experimental evidences for the 1D-HAF state is highly desirable.

Electron paramagnetic resonance (EPR) is a resonance technique that has been broadly used to study low-dimensional quantum antiferromagnets.\cite{Ajiro} Traditionally, temperature dependence of the shift and linewidth of the EPR spectra have been discussed with the spectral moment analysis, i.e., the so-called Kubo-Tomita theory.\cite{Gatteschi} However, this approach was suggested to be inappropriate in the case of 1D-HAFs. Instead, Oshikawa and Affleck  employed the sine-Gordon quantum-field theory to circumvent some computational difficulties of the quantum spin state in 1D-HAFs and to predict the temperature dependence of the EPR linewidth for different magnetic anisotropies.\cite{O-A1,O-A2} Experimental tests on model systems showed an excellent  agreement with this theory.\cite{Zvyagin, Mirta}   Interestingly, exact low-energy solutions for EPR spectra in terms of TLL exist in the spin channel only for metallic wires\cite{Dora_simon} and are still missing for the case of insulating spin-chain systems.        
First EPR studies of CsO$_2$ single crystals were reported already back in the 1970's by K\"{a}nzig,\cite{Koenzig} but clearly the experimental data need to be reinterpreted in light of the new findings. Therefore, we decided to remeasure the X-band EPR spectra and to test the data against the Oshikawa-Affleck and the TLL theories. An  excellent agreement is found with both theoretical models below $T_{\rm{s1}}$, thus providing an independent support for the 1D-HAF state in CsO$_2$. Therefore, CsO$_2$ indeed represents a rare example of the $p-$orbital one-dimensional quantum antiferromagnet.

\section{Experimental methods}

The CsO$_2$ powder sample came from the same batch as the one used in our previous NMR study where the preparation was described in detail.\cite{Martin_CsO2}
Freshly destilled Cs metal was oxidized with $^{17}$O$_2$ gas. To homogenize the yellow product of Cs$^{17}$O$_2$, several cycles of grinding under a controlled argon atmosphere and temperature annealing in natural O$_2$ was processed.
The sample was enriched by the $^{17}$O isotope to approximately 50\%, according to Raman spectroscopy. Laboratory X-ray diffraction at room temperature confirmed high purity of the  sample, as all peaks could be indexed to the  tetragonal, $I4/mmm$, structure of CsO$_2$. No  impurity phase could be traced down to the resolution of our XRD equipment.

The magnetic properties of the sample, which were first checked by SQUID magnetometry,\cite{Martin_CsO2} are in good agreement with previous results.\cite{Palstra} In particular, an inflection in the magnetic susceptibility curve $\chi(T)$ occurring near 70 K on heating ascribed to the structural/orbital ordering transition and a broad hump near 30 K implying the 1D-HAF state, are clearly visible. A Curie-Weiss analysis of the high-temperature $\chi^{-1}(T)$ data obtained in a field of 0.5 T yields an effective magnetic moment $\mu_{\rm eff}$ of $2.01 \mu_{\rm B}$ (here, $\mu_{\rm B}$ is Bohr magneton) and a Curie-Weiss constant $\theta_{\rm CW} = -7$~K. Similarly as in Ref. \onlinecite{Palstra}, $\mu_{\rm eff}$ is considerably larger than the spin-only value of $\mu_{\rm eff}= 1.73\mu_{\rm B}$  for an $S = 1/2$ system, thus confirming an orbital contribution to $\mu_{\rm eff}$ in the high temperature phase of CsO$_2$. The N\'{e}el  ordering transition below 10 K is mostly obscured by a low-temperature Curie tail due to a small amount of  paramagnetic impurities. However,  a weak feature in $\chi (T)$ around 9 K in data measured at low magnetic field of  0.01 T indicates the transition to the ordered state. 

For the EPR measurements, 23.1 mg of powder sample was sealed under dynamic vacuum in a standard Suprasil quartz tube. 
The X-band EPR experiments were performed on a home-built spectrometer equipped with a Varian E-101 microwave bridge, a Varian TEM104 dual cavity resonator, an Oxford Instruments ESR900 cryostat and an Oxford Instruments ITC503 temperature controller.  The temperature stability was better than $\pm 0.05$ K at all temperatures. The EPR spectra were measured on cooling.

\section{Results and discussion}

\begin{figure}[htbp]
  \includegraphics[width=\linewidth]{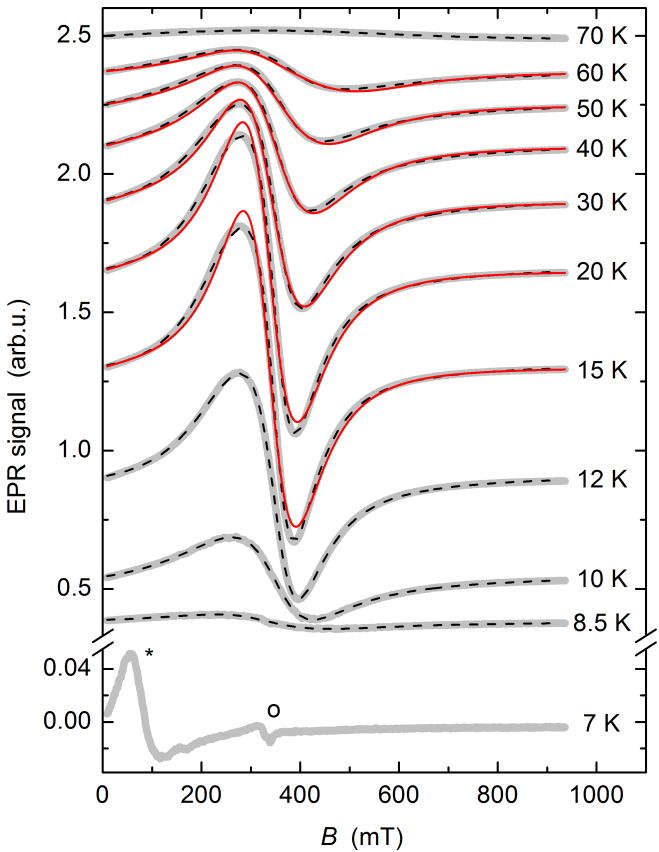}
  \caption[]
  {(color online) The temperature evolution of the X-band EPR spectra in CsO$_2$ powder (grey lines). Dashed black lines are fits with the uniaxial $g$-factor and linewidth anisotropy ($g_{xx} = g_{yy} \neq g_{zz}$ and $\Delta B_{x}=\Delta B_{y}=\Delta B_{z}/2$). Solid red lines are fits with Eq. (\ref{TLL_spekter}) to the TLL spectrum. Below the N\'eel temperature $T_{\rm N}=8.3$~K the paramagnetic EPR signal completely disappears and antiferromagnetic resonance (marked by *) is observed at around 80 mT at $ 7$~K. At $7$~K we observe also an extremely weak impurity line at $g\approx 2.06$ (marked by o).}
   \label{slika_spektri_cezij}
\end{figure}

{\em Temperature evolution of the X-band EPR spectra.}
The CsO$_2$ powder sample shows no detectable X-band EPR signal at room temperature. On cooling, a very broad line starts to appear below $\sim70$~K and becomes very pronounced below $T_{\rm{s1}}$ (Fig. \ref{slika_spektri_cezij}). For temperatures close to $T_{\rm{s1}}$, the EPR spectrum is shifted towards higher fields relative to $g\approx 2$,   thus indicating an orbital angular momentum contribution in the high-temperature orbitally-disordered  phase and consistent with the large high-temperature $\mu_{\rm eff}$ extracted from the SQUID data.\cite{Palstra, Martin_CsO2} However, as temperature decreases well below $T_{\rm{s1}}$, the EPR signal narrows and shifts to $g\approx 2$ resonance-field range. The powder spectra are well  fitted with the uniaxial $g-$factor anisotropy, i.e., $g_{xx} = g_{yy} \neq g_{zz}$, convoluted with a Lorentzian lineshape broadening whose linewidth also has the uniaxial anisotropy, i.e., $\Delta B_{x}=\Delta B_{y}\neq\Delta B_{z}$. At $T= 15$~ K, an unconstrained fit converges to $g_{xx}=g_{yy}= 1.8886$ and $g_{zz}=2.6963$, which  agree with the literature data obtained on single crystals.\cite{Koenzig} Moreover, since we consistently find $\Delta B_x\approx \Delta B_z /2$ at all temperatures, we next constrain $2\Delta B_{x}=\Delta B_{z}\equiv \Delta B$ to find that $\Delta B$ nearly linearly decreases with decreasing temperature  between 60 and $\sim 20$~K. The EPR spectrum is the narrowest at $T=15$~K where $\Delta B = 249$~mT. On further cooling, the signal broadens again as the N\'eel temperature $T_{\rm{N}}=8.3$~K is approached from above before it completely disappears below $T_{\rm{N}}$ and is replaced by another slightly asymmetric signal shifted to significantly lower fields. For example, at $T=7$~K this new resonance is centered at around 80 mT. Since this signal is found only below $T_{\rm{N}}$, we tentatively assign it to one of the antiferromagnetic resonance modes. At lowest temperatures we notice another extremely weak peak at $g\approx 2.06$, which we attribute either to  impurity O$_2^{-}$ centers or to point defects in the antiferromagnetically ordered magnetic lattice.           

\begin{figure}[htbp]
  \includegraphics[width=\linewidth]{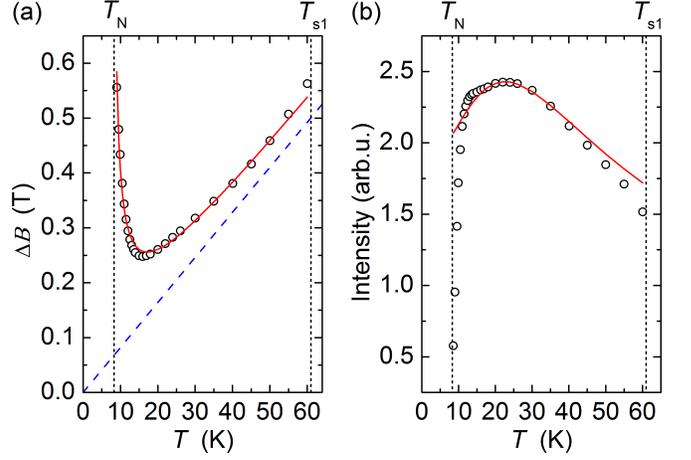}
  \caption[]
   {(color online) (a) Temperature dependence of the X-band EPR linewidth as obtained from the lineshape fits of CsO$_2$ spectra (open black circles). Solid red line represents fits to Eqs. (\ref{OA_fit}) or (\ref{TLL_fit}), which are undistinguishable. Dashed blue line shows the contribution of the 1D-HAF phase to the total EPR linewidth.  (b) Temperature dependence of the X-band EPR signal intensity of CsO$_2$ (open black circles). Solid red line is fit to the static spin susceptibility of the 1D-HAF\cite{Feyer} with $J=35$~K.  Dotted vertical black lines mark the N\'eel ($T_{\rm N}= 8.3$~K) and orbital-ordering temperatures ($T_{\rm s1}= 61$~K).}
   \label{slika_sirina}
\end{figure}

{\em Linewidth analysis with the Oshikawa-Affleck  theory.}
As recent structural and NMR investigations of CsO$_2$ interpreted the phase that develops below $T_{\rm{s1}}$ as a 1D-HAF phase,\cite{Palstra, Martin_CsO2} we proceed to discuss the measured EPR spectra (Fig. \ref{slika_spektri_cezij}) in terms of the one-dimensional magnetism.  
We start with the Oshikawa-Affleck theoretical framework of a  low-temperature EPR response in $S=1/2$ one-dimensional Heisenberg antiferromagnets, which predicts distinct temperature dependences of EPR linewidth for the case of anisotropic exchange broadening, where $\Delta B \propto T$, as opposed to the case of staggered-field broadening, where $\Delta B \propto (B/T)^2$.\cite{O-A2} Since we find that $\Delta B$ nearly linearly increases with increasing temperature between 15~K and $T_{\rm{s1}}$ (Fig. \ref{slika_sirina}a), we conclude that the main anisotropic interaction that broadens EPR spectra is the symmetric anisotropic exchange interaction. A rapid increase in $\Delta B$ on cooling between  15~K and $T_{\rm N}$ is attributed to the critical fluctuations just before the three-dimensional antiferromagnetic order sets in. In the temperature range between $T_{\rm s1}$ and $T_{\rm N}$, the temperature dependence of $\Delta B$ is thus fitted to 
\begin{equation}
\label{OA_fit}
\Delta B = \frac{2\epsilon k_B\delta ^2}{g \mu _B \pi ^3}T + \Delta B_{\rm cf}\left(\frac{T-T_{\rm N}}{T_{\rm N}}\right)^{-\alpha}\quad ,
\end{equation}
which comprises linewidth contributions from the 1D-HAF phase with symmetric anisotropic exchange\cite{O-A2} and from the critical fluctuations, respectively. 
Here, $k_B$ is the Boltzman constant. The constant $\epsilon$ amounts to 2 for magnetic field directed along the anisotropy axis and to 1 for the perpendicular case, thus nicely accounting for the experimental observation $\Delta B_x=\Delta B_z/2$. Since we have defined $\Delta B=\Delta B_z$, we use $\epsilon =2$ in Eq. (\ref{OA_fit}).  An excellent fit of the experimental EPR linewidth (Fig. \ref{slika_sirina}a) yields the constant of symmetric anisotropic exchange $\delta = 0.35 J$. Although we cannot determine the sign of $\delta$, because the corresponding linewidth term is proportional to $\delta^2$, we note a significant magnetic anisotropy that is consistent with the NMR finding of the Ising-like exchange-coupling anisotropy. The magnitude of the critical fluctuations is found to be $\Delta B_{\rm cf}=0.12$~T and the critical exponent $\alpha =0.6$.

{\em Analysis of the EPR spectra within the Tomonaga-Luttinger-Liquid framework.}
The above analysis demonstrates that the EPR data of CsO$_2$ may indeed be interpreted with the 1D-HAF model, at least for $15\,  {\rm K} < T < T_{\rm{s1}}$. However, a recent $^{133}$Cs NMR study of CsO$_2$ employed the TLL formalism instead,\cite{Martin_CsO2} so that it is difficult to directly compare the two results. Therefore, for the quantitative comparison with the NMR data  we next test our EPR data also against the TLL model. 

The EPR spectrum $I(\omega,B)$ is given by the imaginary part of the dynamical spin susceptibility for the transverse direction (i.e., perpendicular to the magnetic field), $\chi_{\perp}''(q,\omega,B)$, calculated at the momentum of the excitation $q=0$ and at the   resonant frequency $\omega$, namely $I(\omega,B)={B_\perp^2\over 2\mu_0}\omega \chi_{\perp}''(q=0,\omega,B)V$.\cite{Schlichter} Here, $B_\perp$ is the magnitude of the perpendicular (excitation) microwave field, $\mu_0$ the permeability of the vacuum and $V$ the volume of the sample. We stress at this point that the EPR spectrum provides complementary information to the NMR spin-lattice relaxation rate, which is proportional to $\chi_{\perp}''(q,\omega,B)$ integrated over the Brillouin zone. At low frequencies, the dynamical spin susceptibility has nonzero contributions in the vicinity of the four $Q$ vectors in the Brillouin zone, i.e., $\pi$ and $2\pi m$ for the transverse direction, and $0$ and $\pi (1-2m)$ for the longitudinal direction (i.e., along the magnetic field). Here $m$ is the field-induced magnetization, which is related to the two TLL parameters, i.e., the dimensionless exponent  $K$ and the velocity of spin excitations $u$, via $m=BK/\pi u$. The first and the third of the listed modes are commensurate, whereas the other two modes are field-dependent (through $m$) and thus incommensurate.  Among these, the EPR spectrum is determined by the transverse mode with nonzero contribution close to $q=0$, i.e., by the mode at $Q=2\pi m$. For this mode the dynamical spin susceptibility takes the form~\cite{Giamarchi}
\begin{eqnarray}
\label{TLL_korelacijska}
\chi_{\perp}''(q,\omega,T) &=& \nonumber \\* 
&-&C_3'T^{2\gamma} \Big[ \beta \big(\frac{\gamma + 2}{2} - \frac{i(\omega + u\delta q)}{4\pi T},(-1-\gamma )\big) \nonumber \\*
&&\beta \big(\frac{\gamma}{2} - \frac{i(\omega - u\delta q)}{4\pi T},(1-\gamma)\big) + \nonumber \\*
 &+& \beta \big(\frac{\gamma}{2} - \frac{i(\omega + u\delta q)}{4\pi T},(1-\gamma)\big) \nonumber \\*
 &&\beta \big(\frac{\gamma + 2}{2} - \frac{i(\omega - u\delta q)}{4\pi T},(-1-\gamma )\big)\Big]\, ,
\end{eqnarray}
where $q=Q+\delta q$ and $\delta q$ is small. Here, $C_3'$ is the non-universal amplitude,\cite{Giamarchi}  $\beta$ is the Euler's beta function and $2\gamma = 2K+1/(2K)-2$. To satisfy the EPR resonance condition  $q=0$, we have to set $\delta q=-Q$, which finally leads to 
\begin{eqnarray}
\label{TLL_spekter}
I(\omega, B, T) = &-&I_0 \omega T^{2\gamma }  \textrm{Im}[F(2 + \gamma,k_1)F(\gamma,k_2) + \nonumber \\*
 &+& F(2 + \gamma,k_2)F(\gamma,k_1)].
\end{eqnarray}
The signal-intensity prefactor $I_0$ is proportional to the electronic static spin susceptibility $\chi_0$, to $B_\perp$ and $V$. The arguments $k_{1,2} =\left( \hbar \omega \mp 2g\mu _B KB\right)/\left(2\pi k_B T\right)$ and the function $F(x,y)= \beta [\left(x-iy\right)/2,1-x]$. We stress that Eq. (\ref{TLL_spekter}) is analogous to the one calculated for the spin channel of the correlated one-dimensional metals, e.g., carbon nanotubes.\cite{Dora_simon} Fitting all the experimental X-band EPR spectra (Fig. \ref{slika_spektri_cezij}) in the temperature range $15\, {\rm K}\leq T\leq 60\, {\rm K}$ to $dI(\omega, B, T)/dB$ with  $I_0$ and $K$ as the only free parameters  produces excellent lineshape fits (solid red lines in Fig. \ref{slika_spektri_cezij}) yielding $K = 0.48$. We note that large broadening of the EPR spectra is found as soon as $K$ deviates significantly from $1/2$, which is again reminiscent of the case of interacting itinerant electrons in one dimension.\cite{Dora_simon}

The temperature dependence of the EPR signal intensity is shown in Fig. \ref{slika_sirina}b. The intensity shows a broad maximum at $T_{\rm max}\approx 22$~K and is rapidly suppressed upon approaching the N\'eel temperature from above. We note that $T_{\rm max}$ matches the temperatures of the maxima in the bulk susceptibility measurements\cite{Palstra} and in the local spin susceptibility probed by $^{133}$Cs NMR.\cite{Martin_CsO2} The best fit of the EPR signal intensity to the 1D-HAF spin susceptibility\cite{Feyer} in the temperature range from 60~K to 15~K is found for $J=35$~K, which is close to the corresponding $J=40$~K extracted from the magnetic susceptibility\cite{Palstra} and $^{133}$Cs NMR data.\cite{Martin_CsO2}    The deviations of the fit from the data points close to $T_{\rm s1}$ are expected due to a large lineshape broadening and lineshift in this temperature range (Fig. \ref{slika_spektri_cezij}), implying underestimated data values, whereas the discrepancies below $\sim 15$~K are due to the onset of three-dimensional antiferromagnetic correlations.
The temperature dependence of the EPR signal intensity and the magnitude of the extracted $J$ thus fully corroborate  the proposed TLL state below $T_{\rm s1}$.

Finally, since in our case  $\gamma \ll 1$ and $k_BT \gg \hbar \omega, g\mu_BB$, we can by using arguments of Ref. \onlinecite{Dora_simon} simplify Eq. (\ref{TLL_spekter}) to a sum of two Lorentzian components, one centered at the positive and the other one at the negative resonance field, with the temperature-dependent linewidth 
\begin{equation}
\label{TLL_fit}
\Delta B _{\rm{TLL}} = \frac{4\pi\gamma k_B}{g \mu _B }T  .
\end{equation}
Not surprisingly, in this limit the temperature dependence of the EPR linewidth  takes, in terms of the temperature dependence, the same form  as Eq. (\ref{OA_fit}). Therefore, fitting the experimental linewidth to 
\begin{equation}
\label{TLL_fit}
\Delta B  = \Delta B_{\rm{TLL}} + \Delta B_{\rm cf}\left(\frac{T-T_N}{T_N}\right)^{-\alpha}\quad ,
\end{equation}
and using the same parameters for critical fluctuations $\Delta B_{\rm cf}$ and $\alpha$ as above, we find $K = 0.48$ consistent with the value of $K$ obtained from the EPR lineshape fits.
{\red To solve the problem of antiferromagnetically coupled $S =1/2$~ spins in one dimension, both the Oshikawa-Affleck and the TLL theories bosonize the spin operators and then use field-theory approaches. However, to calculate the EPR response, the older Oshikawa-Affleck theory\cite{O-A1, O-A2} calculates the EPR width and shift perturbatively. The exact solutions for the dynamic spin susceptibilities were published only later\cite{Giamarchi} and here we take a full advantage of these newly made discoveries to calculate exactly the EPR spectrum. Given the same starting point for the  Oshikawa-Affleck and the TLL approaches, it is thus not surprising that both theories equally well describe the EPR spectra in the parameter range relevant for CsO$_2$. }

The most important finding of this work is thus that the temperature dependence of the EPR lineshape,  the EPR linewidth, and the EPR signal intensity fully comply with the interpretation of the orbitally-ordered phase of  CsO$_2$ in terms of TLL framework, with $K$ consistently converging to the value close to but yet slightly smaller than $1/2$. The extracted $K$ {\red and the symmetric anisotropy-exchange $\delta$}  places CsO$_2$ on the Ising-anisotropy side of the phase diagram. Although this conclusion in general agrees with the previous analysis of the $^{133}$Cs NMR data\cite{Martin_CsO2}, we note that $K\approx 1/2$ derived from the present EPR data is notably  larger than $K\approx 1/4$ extracted from the high-field ($2.35 < B < 9.4$~T) NMR experiments on the same sample. This discrepancy remains puzzling. However, since EPR and NMR probe the low-energy part of $\chi_{\perp}''(q,\omega)$ in a slightly different way, the features in the excitation spectrum arising from the effects of interchain couplings may echo differently in EPR and NMR data. 
{\red The origin of the symmetric anisotropic exchange is unclear at the moment, but most likely it directly reflects the fact that the Cs$^+$ ions, characterized by a significant spin-orbit coupling constant, are bridging the superexchange between nearest neighboring O$_2^-$ units along the $b-$axis.}

\section{Conclusions}

In conclusion, a systematic analysis of the X-band EPR spectra in the low-temperature orbitally-ordered phase of CsO$_2$ agrees with the proposed 1D-HAF state. Using either the Oshikawa-Affleck or the TLL theoretical frameworks, we find a systematic agreement with the experimental EPR lineshape, linewidth and signal intensity, thus allowing us a quantitative determination of the anisotropy of the symmetric exchange interaction or, alternatively, of the TLL exponent. Therefore, our results indeed comply with CsO$_2$ as a  $p-$orbital system showing a quantum antiferromagnetic state where orbital ordering is pivotal in the formation of $S=1/2$ spin chains. Since similar orbital-ordering physics is ubiquitous to the whole family of {\em A}O$_2$ superoxide compounds\cite{Jansen_rev, Sarma,Solo,Kim,Solo2,Oles,Palstra,Palstra2,Martin_CsO2} and {\em A}$_4$O$_6$\cite{Jansen,Arcon_Cs4O6} sesquioxide compounds, it would be interesting to test in the future the low-dimensional quantum magnetism also in these systems.

\begin{acknowledgments}
D.A. and C.F. acknowledge the financial support by the European
Union FP7-NMP-2011-EU-Japan project LEMSUPER
under Contract No. 283214. 
We thank Walter Schnelle and Ralf Koban for performing magnetization measurements.
\end{acknowledgments}

%
%
%
%
%
%
%
%

\end{document}